\documentclass[aps,prl,amsmath,amssymb,reprint,superscriptaddress]{revtex4-1}
\usepackage{graphicx}
\usepackage{amsmath}
\usepackage{natbib}
\usepackage{dcolumn}
\usepackage{bm}
\usepackage{bbold}
\usepackage{color}
\usepackage{amsfonts}
\usepackage{amssymb}
\usepackage{mathrsfs}
\usepackage{tabularx}
\usepackage{braket}
\usepackage{mathtools}
\usepackage{soul}			

\usepackage{caption}
\usepackage{subcaption}

\begin{document} 
\title{Superstrong Coupling of a Microwave Cavity to YIG Magnons}

\author{Nikita Kostylev}
\affiliation{ARC Centre of Excellence for Engineered Quantum Systems, School of Physics, University of Western Australia, 35 Stirling Highway, Crawley WA 6009, Australia}

\author{Maxim Goryachev}
\affiliation{ARC Centre of Excellence for Engineered Quantum Systems, School of Physics, University of Western Australia, 35 Stirling Highway, Crawley WA 6009, Australia}

\author{Michael E. Tobar}
\email{michael.tobar@uwa.edu.au}
\affiliation{ARC Centre of Excellence for Engineered Quantum Systems, School of Physics, University of Western Australia, 35 Stirling Highway, Crawley WA 6009, Australia}

\date{\today}


\begin{abstract}
Multiple-post reentrant 3D lumped cavity modes have been realized to design the concept of discrete Whispering Gallery and Fabry-P\'{e}rot-like Modes for multimode microwave Quantum Electrodynamics experiments. 
Using a magnon spin-wave resonance of a submillimeter-sized Yttrium-Iron-Garnet sphere at milliKelvin temperatures and a four-post cavity, we demonstrate the ultra-strong coupling regime between discrete Whispering Gallery Modes and a magnon resonance with strength of 1.84 GHz. 
By increasing the number of posts to eight and arranging them in a D$_4$ symmetry pattern, we expand the mode structure to that of a discrete Fabry-P\'{e}rot cavity and \textcolor{black}{modify} the Free Spectral Range (FSR). We reach the superstrong coupling regime, where spin-photon coupling strength is larger than FSR, with coupling strength in the 1.1 to 1.5 GHz range.
\end{abstract}

\maketitle

Cavity Quantum Electrodynamics (QED) is a conceptual paradigm dealing with light-matter interaction at the quantum level that has been investigated in a number of various systems. 
There is a broad range of various problems that have to be solved by cavity QED including generation of nonclassical states\cite{Brattke:2001aa}, quantum memory\cite{Maitre:1997aa}, quantum frequency conversion\cite{PhysRevLett.92.247902,PhysRevLett.109.130503}, etc. 
For many of these applications, it is important to combine advantages of different approaches to QED in a Hybrid Quantum System (HQS)\cite{Xiang:2013aa,Kurizki:2015}. For example, combination of nonlinear properties of superconducting circuits based on Josephson Junction and large electron\cite{PhysRevLett.102.083602} or nuclear-spin\cite{Abdurakhimov:2015} ensembles can be used for new quantum protocols without single spin manipulation and is investigated in many physical implementations\cite{PhysRevLett.105.140503,PhysRevLett.107.060502,PhysRevLett.105.140502,Zhu:2011aa,Probst:2013zg}.

In the process of HQS design, it is vital to be able to engineer photon modes by continuous adjustment of system parameters without reinventing a new cavity. It is important to have a single platform that can provide a broad range of spectra required for each particular purpose. Moreover, in order to achieve strong coupling with other elements of HQS, such a platform should guarantee reconfigurable high space localisation of both electrical and magnetic fields to achieve sufficient filling factors. Finally, such cavities are required to be adjustable {\it in-situ} in the wide range preferably at high speed rate. 
These features are lacking for traditional 3D cavities such as box resonators and microwave Whispering Gallery Mode resonators. Having only one or two free parameters to control, these platforms can be hard to modify for a particular set of requirements in terms of field patterns, spectra and tunability without significant change of their structure.

{All the described requirements are met by constructing designs based on the recently proposed multi-post reentrant cavity\cite{Goryachev:2015aa,patent2014} that is based on a known 3D closed resonator with a central post gap\cite{reen0,reen2}.} For this platform, it has been demonstrated that by an {\it a priori} rearrangement of the post, one can easily engineer the device resonance frequencies and field patterns to achieve high frequency and space localisation\cite{Goryachev:2015aa} that guarantees strong coupling regimes\cite{Goryachev:2014ab,diamond}. On the other hand, due to high localisation of electric field in the post gaps, such cavities appear to be highly tunable by mechanical actuators that outperform any kind of magnetic field tuning in terms of speed\cite{reen1}.

\textcolor{black}{In this work, we use some of the discussed capabilities of the reentrant cavity platform in order to reach a new cavity QED interaction regime: superstrong coupling.} This name refers to a regime for which the coupling strength $g$ exceeds not only the spin ensemble $\Gamma$ and cavity $\delta$ loss rates, but also the free spectral range $\omega_{\text{FSR}}$\cite{Meiser:2006aa,Yu2009ssc,Wu2009ssc}. {It has to be noted that a so-called ultrastrong coupling regime, characterised by coupling strengths being comparable to mode frequency\cite{Ciuti2006usc,Niemczyk2010usc,Askenazi2014usc}, has been reached in other works \cite{Zhang:2014,Goryachev:2014ab}. However to achieve superstrong coupling in a QED cavity at microwave frequency, it must not only provide the high filling factor to maximise the coupling strength but allow one to arrange the cavity microwave modes with the desired frequency separation.} Obviously, these goals are hardly achievable with traditional cavities since they usually do not have enough degrees of freedom to control these parameters. On the other hand, the multi-post reentrant cavity gives the option to arrange the post in any suitable way that provides sufficient control over the cavity spectra and field patterns simultaneously.
As for the spin ensemble, we choose a magnon resonance of an Yttrium iron garnet (Y$_3$Fe$_5$O$_{12}$ or YIG)\cite{lvov}. These ferrimagnetic systems recently became a popular subject of study\cite{bhoi:2014,maksymov:2015ri,Maksymov:2015br,Cao:2015,Bai:2015}, as they provide high coupling strengths and low spin losses due to high concentration and ordering of Fe ions and low coupling to phonon modes. In this work, we use single crystal YIG spheres, which have also drawn considerable attention \cite{PhysRevLett.104.077202,PhysRevB.82.104413,PhysRevLett.113.083603,lambert:2015,Goryachev:2014ab}. 

In order to achieve the superstrong coupling regime, we design cavities exhibiting at least two resonances separated by $\omega_{FSR}$. Because each post represents a harmonic oscillator, the total system exhibits the number of resonances equal to the number of posts $N$. Each cavity mode is characterised by a unique combination of currents at the same instance of time and, as a result, the magnetic field pattern. So, to couple the cavity to spin modes in a crystal, posts may be arranged to maximise the field in a small volume. Using this property,
it has been demonstrated \cite{Goryachev:2014ab,diamond} how a two-post cavity exhibits dark ($\uparrow\uparrow$ currents) and bright ($\uparrow\downarrow$ currents) modes with maximum and minimum magnetic field in small crystal samples between two posts. Although, in order to achieve the superstrong coupling, it is required to have more cavity modes with large spin-photon coupling. This may be achieved by increasing the number of posts arranged in patterns of certain symmetries to control the free spectral range. 

The experimental setup used in this work is similar to previous experiments\cite{Goryachev:2014ab,diamond}: \textcolor{black}{Reentrant cavities with straight excitation antennas are thermalised to a 20mK stage of a dilution refrigerator inside a superconducting magnet\cite{PhysRevB.88.224426,quartzG}. The excitation signal is attenuated by 40dB at various stages of the cryocooler, whereas the output signal is amplified by a cold low noise amplifier.}

\textcolor{black}{The cavities are fabricated of Oxygen Free Copper. They are 10mm in diameter and contain posts 3.4mm tall. The dielectric gap between the posts and the lid is 0.1mm. The spherical YIG  sample is positioned between posts at the centre of the cavity and is held in place by a teflon mount. As a non-superconducting, relatively high-loss material has been used, quality factors $Q$ of modes are not expected to be large. High $Q$s are not required in this experiment due to the very high coupling strength. They can be improved by using silver or niobium cavities, optimising the geometric factor $G_c$ of the system or adjusting positions and dimensions of the posts\textcolor{black}{\cite{Goryachev:2014ab}}. 
 }

\begin{figure}[!ht]
	\includegraphics[width=0.45\textwidth]{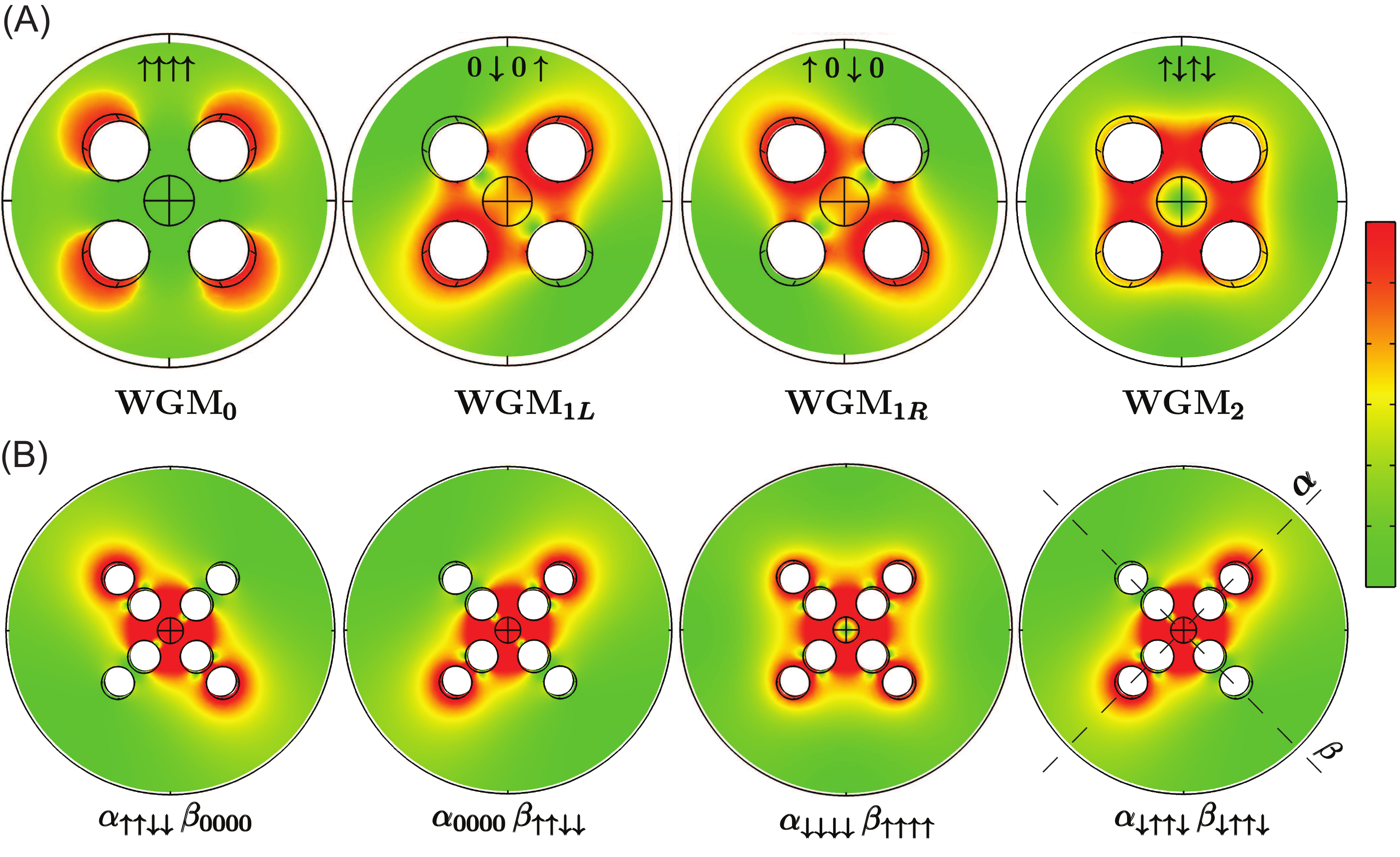}
	\caption{Magnetic field distribution at the equator of YIG sphere inside the $N=4$ (A) and $N=8$ (B) post cavities. The modes are shown as a function of increasing frequency (from left to right). Only four modes of interest out of eight are shown for the (B) graph. }
	\label{comsol}
\end{figure}

The first cavity of $N=4$ with D$_4$ symmetry demonstrates four modes with the following combination of currents at the same moment: $\uparrow\uparrow\uparrow\uparrow$-$\uparrow\!0\!\downarrow\!0$-$0\!\downarrow\!0\!\uparrow$-$\uparrow\downarrow\uparrow\downarrow$ where $0$ denotes the post with no current. Fig.~\ref{comsol}, (A), obtained by finite-element modelling, demonstrates \textcolor{black}{the strength of magnetic field at the equator of YIG sphere, perpendicular to cavity posts}. In an ideal case, the second and the third modes are degenerate in frequency because one is $\pi/2$ rotation of another. They represent a degenerate mode pair, similar to so-called Whispering Gallery Mode doublet, a pair of sine and cosine waves, since the mode structure may be understood as a discrete WGM system. This particular doublet represents a WGM of the order $n=2$, since it has two nodes. It has to be pointed out that for each resonance of the doublet all four posts are involved in oscillation even though two of them are not illuminated at some instance of time. In an actual experiment, the D$_4$ symmetry is broken leading to lifting of the mode degeneracy with the frequency splitting depending on the cavity imperfections. This type of an avoided crossing is typical to spin-photon interaction in the cavity with time-reversal symmetry breaking\cite{PhysRevA.89.013810, Goryachev:2014aa} where WGM doublets are formed by travelling waves. \textcolor{black}{In such a situation the cavity doublet pair is coupled together. However, the coupling to the magnon modes is asymmetric with one of the cavity modes hybridizing with the magnon mode in the ultra-strong coupling regime, while the other cavity mode is nearly uncoupled from the magnon mode.}

{The second cavity with $N=8$ with D$_4$ symmetry may be regarded as two perpendicular discrete Fabry-P\'erot systems made of four posts each. 
It is important to underline that the first and the second modes of this structure, $\alpha_{\uparrow\uparrow\downarrow\downarrow} \, \beta_{0000}$ and $\alpha_{0000} \, \beta_{\uparrow\uparrow\downarrow\downarrow}$ respectively (shown in Fig.~\ref{comsol}, (B)) are modes which have a field structure similar to that of two linear Fabry-P\'erot resonators $\alpha$ and $\beta$ and are formed by two chains of four posts\cite{Goryachev:2015aa}.} The indeces denote the direction of currents in the posts. These two modes may be classified as one dimensional modes of order one. The simulated \textcolor{black}{magnetic} field profile for this cavity is shown in Fig.~\ref{comsol}, (B). In this regards, the next mode $\alpha_{\downarrow\downarrow\downarrow\downarrow} \, \beta_{\uparrow\uparrow\uparrow\uparrow}$ can be understood as a combination of zero-order modes for each of the linear resonators.
Similar to the case of $N=4$, in an actual experiment, resonance frequencies of these two cavities are split as the symmetry is unavoidably broken. 
\textcolor{black}{It has to be noted that there exist additional 3 higher- and 1 lower-frequency modes, which are not of interest for this experiment. A more detailed discussion on modes of discrete Fabry-P\'erot cavities is available in another work\cite{goryachev2015cavity}}.

\begin{figure}[!ht]
	\includegraphics[width=0.42\textwidth]{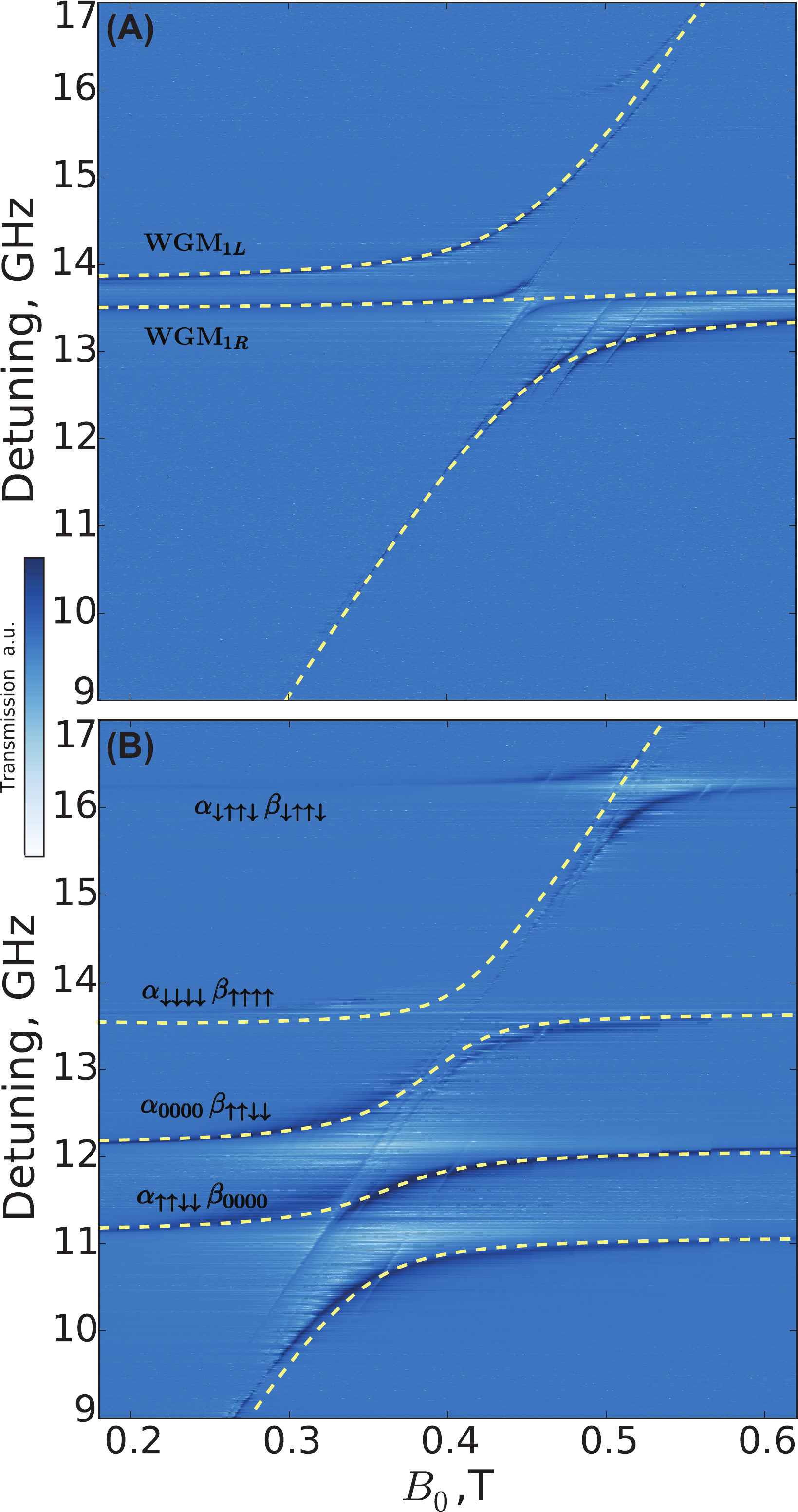}
	\caption{Transmission through $N=4$ (A) and $N=8$ (B) post cavities as function of the driving frequency and the external magnetic field. The dashed curves are theoretical predictions for the system eigenfrequencies.}
	\label{interaction}
\end{figure}
\begin{figure}[!ht]
	\includegraphics[width=0.5\textwidth]{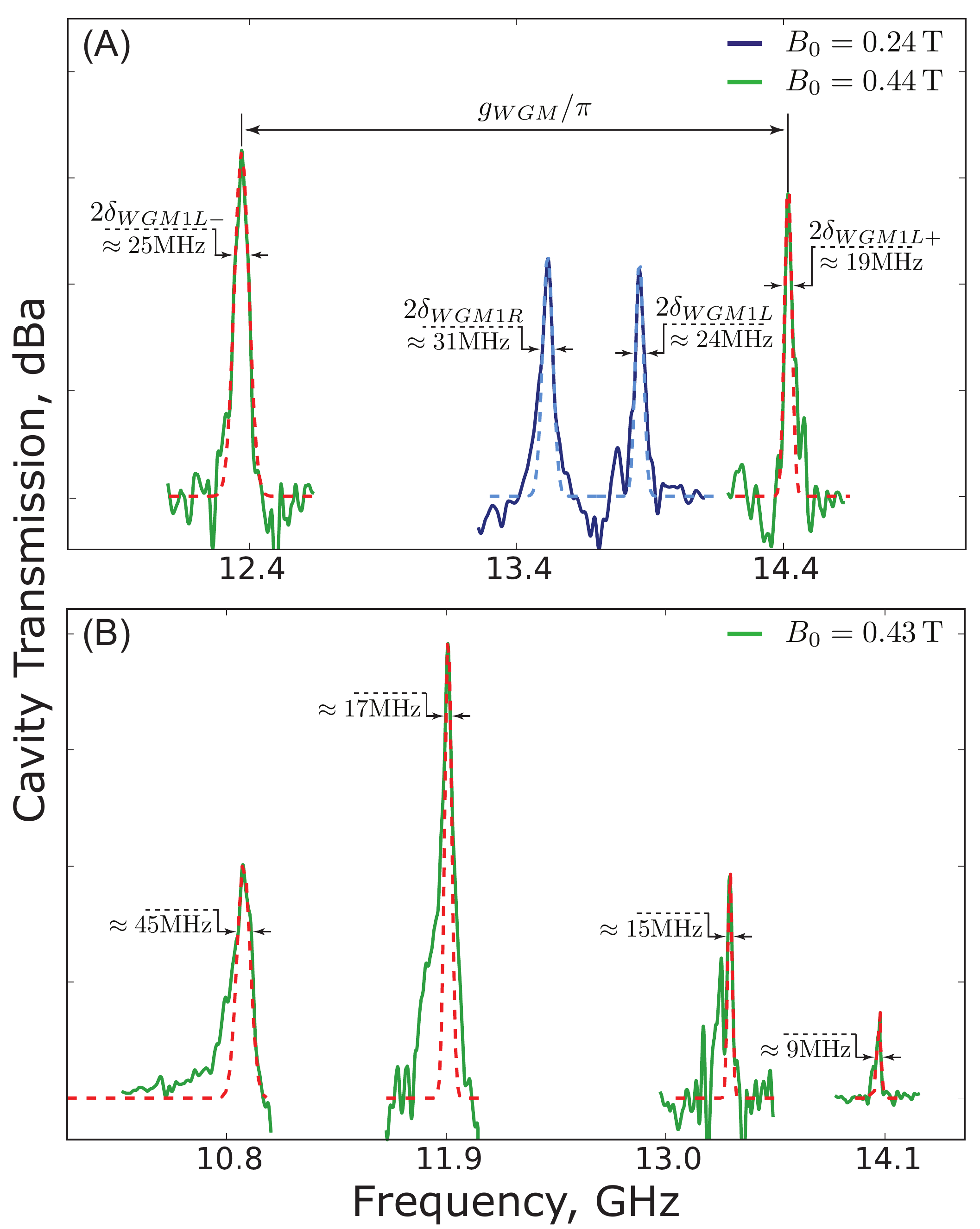}
	\caption{Transmission through $N=4$ (A) and $N=8$ (B) post cavities as function of the driving frequency for a chosen external magnetic field. Plot (A) shows the frequency response of the interaction between the WGM$_{1L}$ and WGM$_{1R}$ cavity modes and the magnon mode. Plot (B) shows the resonant frequency response of the 8-post cavity, demonstrating superstrong coupling. Dashed curves represent Lorentzian fits to the data. Linewidths are given for the case when the magnon resonance is tuned onto the cavity mode. }
	\label{bcut}
\end{figure}

The experimental results of magnon-photon interaction for both cavities are shown in  Fig.~\ref{interaction}. \textcolor{black}{ Fig.~\ref{interaction}(A) corresponds to the measurement of $N=4$ cavity, loaded with a 0.8 mm diameter YIG sphere, and demonstrates an Avoided-Level Crossing (ALC) between one of the cavity doublet modes and a magnon resonance.} The other doublet mode does not interact with the YIG sphere for symmetry reasons \cite{Goryachev:2014ab}. For this cavity, the system Hamiltonian relates annihilation (creation) operators $a_R$ ($a_R^\dagger$) and $a_L$ ($a_L^\dagger$) of photon modes WGM$_{1R}$ and WGM$_{1L}$ (shown in Fig.~\ref{comsol}, (A)) to $b$ ($b^\dagger$), that is, annihilation (creation) operators of the \textcolor{black}{uncoupled} magnon mode, in units where $\hbar =$ 1:
\begin{multline}
	\label{B006SFa}
	\displaystyle  H_{N=4} = \omega_c(a_R^\dagger a_R+a_L^\dagger a_L)+G_{RL}(a_R^\dagger + a_R) (a_L^\dagger + a_L) \\
	\displaystyle + \omega_{m} b^\dagger b  + g(a_R^\dagger + a_R) (b^\dagger + b).
\end{multline}
Here $\omega_c$ is the cavity angular frequency, $G_{RL}$ is the asymmetry induced coupling between photon doublet modes,  $\omega_{m}$ is the field controllable angular frequency of the magnon mode, and $g$ is the photon-magnon coupling strength. Note that here we ignore all higher order magnon modes. Fig.~\ref{interaction}, (A) demonstrates fitting of the experimentally measured resonance frequencies to the three mode model~(\ref{B006SFa}). The fit reveals the following values for the model: $\omega_c/(2\pi) = $ 13.65 GHz, $G_{RL}/(2\pi) = $ 155 MHz and $g = $ 1.84 GHz. \textcolor{black}{With the spin density in YIG of $2\times10^{22}$cm$^{-3}$\cite{Huebl2013high}, the filling factor $\xi$ is estimated as $1.5\times10^{-2}$, which is very high and in good agreement with finite-element modelling.} The profile of the modes (in transmission) measured as a function of frequency for the fields $B_0 = $ 0.24 T and 0.44 T  is shown in Fig.~\ref{bcut}, (A). The cavity linewidths \textcolor{black}{away from the magnon resonance} for the WGM$_{1R}$ and WGM$_{1L}$ modes have been measured as 14 MHz and 22 MHz, corresponding to $Q$ factors of 969 and 643 respectively. Magnon linewidth has been found to be on the order of 1 MHz,  in agreement with previous work \cite{Goryachev:2014ab}.

Fig.~\ref{interaction}, (B) shows the magnetic field response for the case of $N = 8$, where the magnon resonance line exhibits a number of ALCs with cavity modes. \textcolor{black}{A 1.0 mm diameter optically polished YIG sphere was used for this experiment.} The Hamiltonian for this system, ignoring higher order cavity and magnon modes, is written as follows:
\begin{multline}
	\label{B007SFa}
	\displaystyle  H_{N=8} = \omega_{c1}a_\alpha^\dagger a_\alpha + \omega_{c2} a_\beta^\dagger a_\beta + \omega_{c3} a_{\alpha\beta}^\dagger a_{\alpha\beta}\\
	\displaystyle  + \omega_m b^\dagger b
	\displaystyle + \sum\limits_{i} g_i(a_i + a_i^\dagger) (b + b^\dagger),
\end{multline}
where $i\in\{\alpha,\beta,\alpha\beta\}$, $a_\alpha^\dagger$ ($a_\alpha$) and $a_\beta^\dagger$ ($a_\beta$) are creation (annihilation) operators for cavity modes $\alpha_{\uparrow \uparrow \downarrow \downarrow} \, \beta_{0000}$ and $\alpha_{0000} \, \beta_{\uparrow \uparrow \downarrow \downarrow}$, with angular frequencies $\omega_{c1}$ and $\omega_{c2}$. $a_{\alpha\beta}^\dagger$ ($a_{\alpha\beta}$) are creation (annihilation) operators for the mode $\alpha_{\downarrow\downarrow\downarrow\downarrow} \, \beta_{\uparrow\uparrow\uparrow\uparrow}$ with angular frequency $\omega_{c3}$. As in (\ref{B006SFa}), $b^\dagger$ ($b$) describe the creation (annihilation) of the magnon mode with field-dependent frequency of precession $\omega_m$. The parameter $g_i$ determines the strength of the photon-magnon coupling for the $i$-th mode.  The fit of this $N=8$ model to experimental data (Fig.~\ref{interaction}, (B)) gives the following values for the couplings: $\omega_{c1}/(2\pi) = $ 11.20 GHz, $\omega_{c2}/(2\pi) = $ 12.20 GHz, $\omega_{c3}/(2\pi) = $ 13.65 GHz, and $g_i/\pi = $ (1.18 GHz, 1.46 GHz, 1.37 GHz). \textcolor{black}{These values correspond to $\xi\approx1\times10^{-2}$. Such a large filling factor is expected for this type of cavity and agrees well with numerical simulations. The FSR between $\omega_{c1}/(2\pi)$ and $\omega_{c2}/(2\pi)$ is 1 Ghz, which is smaller than the corresponding couplings of 1.18 GHz, 1.46 GHz respectively, indicating that the system has reached the superstrong coupling regime.} The profile of the modes (in transmission) measured as a function of frequency for the field $B_0 = $ 0.43 T is shown in Fig.~\ref{bcut}, (B). \textcolor{black}{ Away from the magnon resonance, the}  linewidths of 36 MHz ($Q = $ 314), 15 MHz ($Q = $ 818), 16 MHz ($Q = $ 859) and 63 MHz ($Q = $ 195) have been measured for the cavity modes $\alpha_{\uparrow\uparrow\downarrow\downarrow} \, \beta_{0000}$ , $\alpha_{0000} \, \beta_{\uparrow\uparrow\downarrow\downarrow}$, $\alpha_{\downarrow\downarrow\downarrow\downarrow} \, \beta_{\uparrow\uparrow\uparrow\uparrow}$ and $\alpha_{\downarrow\uparrow\uparrow\downarrow} \, \beta_{\downarrow\uparrow\uparrow\downarrow}$ respectively.

In conclusion, based on a multi-post reentrant cavity platform, we have developed a technique for engineering the cavity frequency response, as well as spacial field distribution, achieving high frequency and space localisation of modes. Such resonators can be made mechanically tuneable due to high degree of localisation of electric field. We have demonstrated how the reentrant cavity platform can be used to generate a desired mode pattern, including WGM doublet modes, and reached a superstrong coupling regime in YIG crystal, where spin-photon coupling strength is larger than $\omega_{FSR}$.

This work was supported by the Australian Research Council Grant No. CE110001013.


%

\end{document}